\documentclass[12pt,aps,pra,amsmath,amssymb,nofootinbib]{revtex4-2}

\usepackage{orcidlink}
\usepackage{tensor}

\hypersetup{colorlinks,citecolor=blue}
\hypersetup{colorlinks=true,linkcolor=magenta,filecolor=magenta,    urlcolor=blue}
\begin{document}

\title{Conservative cosmology in scalar-tensor Herglotz $f(R,T)$ gravity}

\author{Marek Wazny}
\email{marek.wazny@stud.ubbcluj.ro}
\affiliation{Department of Physics, Babeș-Bolyai University, 1 Kogălniceanu Street, Cluj-Napoca 400084, Romania}

\begin{abstract}
The scalar-tensor representation of $f(R,T)$ gravity is extended to incorporate the Herglotz variational principle. The field equations are derived in both the geometric and scalar-tensor frameworks. Although the divergence of the energy-momentum tensor in matter-geometry coupling theories is generally nonvanishing, conservation can be achieved through the introduction of the Herglotz vector. The generalized Friedmann equations in scalar-tensor Herglotz $f(R,T)$ theory are obtained, and a conservative cosmological model is shown to be consistent with late-time observational data. Comparisons with analogous nonconservative models and with the standard $\Lambda$CDM model are also provided.
\end{abstract}

\maketitle
\tableofcontents

\section{Introduction}

In recent decades, modified theories of gravity have attracted growing interest as alternatives to General Relativity (GR), especially in view of the unresolved problems in cosmology such as dark energy and the late-time accelerated expansion of the Universe \cite{Riess_1998, Perlmutter_1999,deBernardis2000,Hanany_2000}. The most widely accepted cosmological model describing these unresolved problems is $\Lambda$CDM ($\Lambda$ cold dark matter). Although $\Lambda$CDM generally agrees with observational data \cite{moresco2020HzTable}, it still retains challenges such as the cosmological constant problem \cite{weinberg2000cosmologicalconstantproblemstalk} and the Hubble tension \cite{Tanabashi,Riess_2019}.

Therefore, another avenue to overcome these challenges has lead to the modification of gravity. Among the broad class of $f(R,\text{Matter})$ theories \cite{Nojiri:2007bt,Mohseni:2009ns,Harko:2018gxr,Wu:2018idg,Haghani:2021fpx,deLimaJunior:2024icz}, $f(R,T)$ gravity has emerged as a compelling framework, wherein the gravitational Lagrangian is promoted from a dependence on the Ricci scalar $R$ to a general function $f(R,T)$ of both curvature scalar and the trace $T$ of the energy-momentum tensor \cite{Harko_2011, Harko:2010mv, ShabaniFarhoudi2014, Chakraborty2013}. The explicit nonminimal coupling between matter and curvature generally induces nonconservation of the energy-momentum tensor, leading to non-geodesic motion \cite{Harko_2011}, although some conservative theories may be established \cite{PhysRevD.107.124005, dosSantos2018}. These features have inspired extensive cosmological studies demonstrating that $f(R,T)$ gravity can reproduce late-time acceleration without invoking a separate dark energy component \cite{ Zaregonbadi2016}.

A method for analyzing modified gravity theories is their reformulation in terms of scalar-tensor representations. Such representations render  more transparent the curvature and matter degrees of freedom separately. In the context of $f(R)$ gravity, the scalar-tensor equivalence with Brans--Dicke-type theories has been thoroughly established \cite{SotiriouFaraoni2010, DeFeliceTsujikawa2010}. Extensions involving curvature-matter couplings may also be cast into scalar-tensor form \cite{Bertolami2007}. More recently, scalar-tensor formulations of $f(R,T)$ gravity have been examined in detail, yielding insights into the thermodynamic interpretation of the nonconservation of the energy-momentum tensor with particle production, and the viability of cosmological solutions \cite{Houndjo2012, Pinto_2022, Bouali_2023}. Models have also been explored in non-minimally coupled theories with conserved energy-momentum tensors \cite{Lobo_2022}, although this heavily constrains the matter scalar. 

In parallel, an independent line of development has advanced the use of nonconservative variational principles in gravitational theories. The Herglotz variational principle, originally formulated to describe classical dissipative systems \cite{Herg1, Herg2}, generalizes the standard action principle by allowing the Lagrangian to depend explicitly on the action itself. The resulting dynamics are governed by differential equations for the action, naturally leading to dissipative terms. Recently, this framework has been applied to cosmological models \cite{Lazo_2017, Lazo_2018, Paiva_2022}, compact-object solutions \cite{PhysRevD.99.124031, PhysRevD.103.044018}, and more general settings \cite{deLeon:2024ztn}.

Motivated by the nonconservation inherent in $f(R,T)$ gravity, the Herglotz variational principle has recently been applied to modified gravity, giving rise to Herglotz-type $f(R,T)$ formulations \cite{Wazny_2025}. It was shown that certain $f(R,T)$ models that were previously deemed cosmologically unviable, such as linear combinations of the form $f(R,T)=R+\alpha T$, can be rehabilitated within the Herglotz variational principle. Additionally, the normally strict behavior of geometry-matter coupling theories to break energy-momentum conservation was shown to be alleviated via the Herglotz vector. This prompts future work exploring the scalar-tensor representation of $f(R,T)$ gravity or, more generally, $f(R,\text{Matter})$ theories within the Herglotz variational framework. Additionally, it could prove useful to apply the Herglotz variational principle to general scalar-tensor theories of Horndeski type \cite{Horndeski1974, Kobayashi_2019}, and investigate the modifications to already established cosmological \cite{CLIFTON20121} and compact object solutions \cite{kobayashi, maselli}.

As such, it will be the goal of this work to extend Herglotz-type $f(R,T)$ gravity into an equivalent scalar-tensor representation and show it provides a cosmological model that is consistent with late-time observational data and the $\Lambda$CDM model. In Section \ref{sec1} an overview of the Herglotz variational principle is established, with specific cases towards classical particle mechanics, field theories, general relativity, and $f(R,T)$ gravity. In Section \ref{sec2} the scalar-tensor approach to Herglotz-type $f(R,T)$ gravity is derived. Field equations are obtained and the equivalence to the geometric approach shown. Furthermore, the cosmology of the theory is explored. In particular, the generalized Friedman equations are derived and a specific form of the scalar potential $U(\varphi,\psi)$ is chosen. Within this specific model the Herglotz vector is used to retain a conservative theory which is then numerical solved for varying initial conditions of the scalar field $\varphi$. Finally, the results are compared to cosmological observations and the standard $\Lambda$CDM model. In Section \ref{conc} final remarks are made.

\section{Herglotz variational principle}\label{sec1}

In this Section we revisit the Herglotz variational principle \cite{Herg1}, an alternative to the usual Lagrangian approach. We begin with its classical, single–particle formulation and then summarize the covariant generalization to field theory as developed by \cite{Lazo_2018}, focusing especially on its application to general relativity (GR). In this context, the appearance of dissipative effects and their implications for energy–momentum conservation will be examined.

\subsection{Generalized Euler--Lagrange equations in classical physics}

The Lagrangian variational method plays a central role in modern theoretical physics. For a mechanical system described by coordinates $q(t)$ and Lagrangian $L(q(t),\dot q(t),t)$, the stationarity of the action yields the familiar Euler–Lagrange equations
\begin{equation}\label{Leom}
\frac{\partial L}{\partial q} - \frac{d}{dt}\frac{\partial L}{\partial\dot{q}} =0 .
\end{equation}
If $L$ has no explicit dependence on time, Eq.~(\ref{Leom}) makes it evident that no first–order derivative terms proportional to $\dot q$ can emerge without introducing explicit time dependence. Consequently, within the standard variational framework, dissipative forces, which in Newtonian mechanics are typically linear in $\dot q$, are difficult to implement.

The Herglotz variational principle resolves this limitation by allowing the Lagrangian to depend on an auxiliary scalar quantity $S$, interpreted as the action itself. Starting from the differential relation
\begin{equation}\label{HerglotzDEaction}
\dot{S} = L(q(t), \dot{q}(t), S(t), t) ,
\end{equation}
one no longer obtains the usual least‐action formulation upon integration. Instead, the variation is performed directly yielding the differential equation
\begin{align}\label{zetaDE}
\delta\dot{S} = \frac{\partial L}{\partial q}\delta q+\frac{\partial L}{\partial\dot{q}}\delta\dot{q} +\frac{\partial L}{\partial S}\delta S .
\end{align}
Imposing the usual stationarity condition $\delta S = 0$ on the boundary leads to the generalized Euler–Lagrange equations \cite{Herg2},
\begin{equation}\label{HerglotzELeqs}
\frac{\partial L}{\partial q}-\frac{d}{dt}\frac{\partial L}{\partial \dot{q}}+\gamma\frac{\partial L}{\partial \dot{q}} = 0 ,
\end{equation}
where 
\begin{equation}
\gamma = \frac{\partial L}{\partial S}.
\end{equation}
This reduces to Eq.~(\ref{Leom}) whenever $\gamma = 0$. This additional term is precisely what allows dissipative dynamics even in the absence of explicit time dependence.

A simple illustration is provided by the one-dimensional Lagrangian with spatial coordinate $x$
\begin{equation}\label{particleL}
L = \frac{1}{2}m\dot{x}^2-U(x) -\frac{\nu}{m}S,
\end{equation}
where $m$ is the mass of the particle, $U(x)$ is a potential, and $\nu$ is the friction coefficient. This system yields the equation of motion
\begin{equation}
m\ddot{x} +\nu\dot{x} = -\frac{dU(x)}{dx} ,
\end{equation}
showing how the dissipative term $\nu\dot x$ arises.

\subsection{The Herglotz variational problem in classical field theories}

A fully covariant formulation of the Herglotz variational principle remained elusive for a considerable time, until it was achieved only recently \cite{ Lazo_2018}. Extending Eq.~(\ref{HerglotzDEaction}) requires promoting all generalized coordinates to fields $\phi \left(\tensor{x}{^\mu} \right)$, and defining an action density field (vector field) $\tensor{s}{^\mu}$, which is differentiable on the domain $\Omega \in \mathbb{R}^n$ with boundary $\partial\Omega\in \mathbb{R}^{n-1}$. The action becomes
\begin{equation}\label{actiondensity}
    S  = \int_\Omega \tensor{\partial}{_\mu} \tensor{s}{^\mu} \,\tensor{d}{^n}x = \int_{\partial\Omega} \tensor{n}{_\mu} \tensor{s}{^\mu}\, \tensor{d}{^{n-1}}x,
\end{equation}
where $n_\mu$ is the unit normal vector to $\partial\Omega$ and the divergence theorem was used. Then, the covariant generalization of Eq. (\ref{HerglotzDEaction}) reads
\begin{equation}
    \tensor{\partial}{_\mu} \tensor{s}{^\mu} = \mathcal{L}\left(\phi\left(\tensor{x}{^\mu}\right), \tensor{\partial}{_\nu} \phi\left(\tensor{x}{^\mu}\right), \tensor{s}{^\mu}, \tensor{x}{^\mu}\right),
\end{equation}
with boundary conditions
\begin{equation}
    \phi(\partial\Omega) = \phi_{\partial\Omega},\quad \phi(\partial\Omega): \partial\Omega \rightarrow \mathbb{R} ,
\end{equation}
where $\mathcal{L}$ is the Lagrangian density. Then, the variation leads to the generalized Euler-Lagrange equations for fields \cite{Lazo_2018}
\begin{equation}\label{HerglotzFieldELeq}
    \frac{\partial \mathcal{L}}{\partial \phi} - \partial_\mu\frac{\partial \mathcal{L}}{\partial\left( \partial_\mu \phi \right)}+ \gamma_\mu \frac{\partial \mathcal{L}}{\partial\left( \partial_\mu \phi \right)} = 0 ,
\end{equation}
where
\begin{equation}
     \gamma_\mu = \frac{\partial \mathcal{L}}{\partial s^\mu} .
\end{equation}
Again, this reduces to the regular Euler-Lagrange equations for fields when $\gamma_\mu=0$. 

An interesting example occurs when considering a canonical scalar field $\phi(x)$ with mass term and Herglotz contribution. The resulting Lagrangian density is
\begin{equation}
    \mathcal{L} = \frac{1}{2}\partial_\mu \phi \partial^\mu \phi  -\frac{1}{2}m^2\phi^2 -\nu_\mu s^\mu,
\end{equation}
for constant vector $\nu_\mu$. The resulting equations of motion become
\begin{equation*}
    (\partial_\mu \partial^\mu +\nu^\mu \partial_\mu +m^2)\phi = 0,
\end{equation*}
representing a damped Klein-Gordon equation. In fact if $\nu^{\mu} = (\alpha,0,0,0) $ this leads to a telegraph style equation with finite propagation speeds in relativistic heat conduction \cite{Jou1993}. This is relevant for example for thermal dynamics in general relativity \cite{doi:10.1098/rspa.2010.0308}.

\subsection{Dissipative gravitational field equations}
The Herglotz variational principle can be extended to general relativity in a straightforward manner but requires some care when obtaining the field equations. Consider an $n$-dimensional manifold $\mathcal{M}$ with subset $\mathcal{V}$ and boundary $\Omega$, then variational setup is formulated as
\begin{align}\label{GRsys}
\begin{split}
\nabla_\mu s^\mu &= \mathcal{L}\left(g_{\alpha\beta}(x^\mu), \partial_\mu g_{\alpha\beta}(x^\mu), s^\mu, x^\mu\right),\\
S(\Omega) &= \int_\Omega d^{n-1}x\sqrt{h} n_\mu s^\mu = \int_\mathcal{V} d^nx \sqrt{-g} ~\nabla_\mu s^\mu ,
\end{split}
\end{align}
where $s^\mu$ is the action density, $h$ and $g$ are the induced and spacetime metric determinants, respectively, and $n_\mu$ is the outward–pointing unit normal.

Introducing the Herglotz field as a closed one form $\lambda = \lambda_\mu dx^\mu$, one can add an effective dissipation term to the typical GR Lagrangian density. The corresponding Lagrangian density is\footnote{The unit convention $c=G=1$ will be used throughout this work unless stated with explicit units.} 
\begin{equation}
\mathcal{L} = R+16\pi\mathcal{L}_m +\lambda_\mu s^\mu ,
\end{equation}
with $R$ the Ricci scalar and $\mathcal{L}_m$ the matter Lagrangian density. Following the derivation in \cite{Paiva_2022}, one obtains for the modified Einstein equations,
\begin{equation}\label{GRHFieldeqs}
G_{\mu \nu} + K_{\mu \nu} = 8\pi T_{\mu\nu} ,
\end{equation}
where $G_{\mu\nu}$ and $T_{\mu\nu}$ are the usual Einstein and energy-momentum tensor and 
\begin{align}
K_{\mu\nu} = \frac{1}{2}\left( \nabla_\mu\lambda_\nu + \nabla_\nu \lambda_\mu\right) -\lambda_\mu \lambda_\nu - g_{\mu\nu}\left(\nabla_\rho \lambda^\rho-\lambda_\rho \lambda^\rho\right),
\end{align}
encodes the dissipative terms.

Taking the divergence of Eq.~(\ref{GRHFieldeqs}) leads to generalized conservation equation
\begin{equation}
8\pi\nabla^\mu T_{\mu\nu} = \nabla^\mu K_{\mu\nu}.
\end{equation}
Thus, the presence of $\lambda_\mu$ generically leads to nonconservation of $T_{\mu\nu}$. Alternatively, requiring a conservative theory forces the constraint $\nabla^\mu K_{\mu\nu}=0$, effectively supplying dynamics for the background field $\lambda_\mu$.

\subsection{Geometrical Herglotz-type $f(R,T)$ gravity}

The Herglotz mechanism may also be incorporated into $f(R,T)$ gravity \cite{Wazny_2025}. In geometric form, the Lagrangian density is
\begin{equation}\label{GeoL}
\mathcal{L} = f(R,T)+\lambda_\mu s^\mu + 16\pi\mathcal{L}_m ,
\end{equation}
where $f(R,T)$ is a well behaved function of the Ricci scalar $R$ and trace of the energy-momentum tensor $T$. The variation of Eq.~(\ref{GRsys}) with respect to the metric tensor yields
\begin{equation}\label{Herglotzeom}
f_{R} R_{\mu \nu} - \frac{1}{2} f g_{\mu\nu} + \left( g_{\mu \nu} \Box - \nabla_{\mu} \nabla_{\nu} \right) f_{R}+H_{\mu \nu} = 8\pi T_{\mu \nu} -T_{\mu \nu} f_{T} - \Theta_{\mu \nu} f_{T},
\end{equation}
where subscripts denote partial differentiation with respect to that variable, $\Box=\nabla^\mu \nabla_\mu$, and
\begin{align}
H_{\mu\nu} &= f_R K_{\mu \nu}+\lambda_\mu \partial_\nu f_R +\lambda_\nu \partial_\mu f_R - 2g_{\mu\nu}\lambda_\rho \partial^\rho f_R , \\
\Theta_{\mu\nu} &= g^{\alpha\beta}\frac{\delta T_{\alpha\beta}}{\delta g^{\mu\nu}} .
\end{align}

Taking the divergence of Eq.~(\ref{Herglotzeom}) and employing the generalized Bianchi identity $[\nabla_\mu,\nabla_\nu]\nabla^\mu f_R + R_{\mu\nu}\nabla^\mu f_R = 0$ yields
\begin{equation}\label{EMdiv}
(8\pi +f_T)\nabla^\mu T_{\mu\nu} = \nabla^\mu H_{\mu\nu}+\frac{1}{2}f_T\nabla_\nu(2\mathcal{L}_m-T) +(T_{\mu\nu}+\Theta_{\mu\nu})\nabla^\mu f_T .
\end{equation}
by expanding $\Theta_{\mu\nu}$ to first order in the metric tensor \cite{Harko_2011}
\begin{equation}
\Theta_{\mu \nu}=-2 T_{\mu \nu} + g_{\mu \nu} \mathcal{L}_{m}.
\end{equation}
Looking at Eq.~(\ref{EMdiv}) $T_{\mu\nu}$ is generally non-conserved, just as in ordinary $f(R,T)$ gravity. Contrary to ordinary $f(R,T)$ however, one can enforce conservation by imposing the condition
\begin{equation}\label{Herglotzdiv}
\nabla^\mu H_{\mu\nu}=\frac{1}{2}f_T\nabla_\nu(T-2\mathcal{L}_m) -(T_{\mu\nu}+\Theta_{\mu\nu})\nabla^\mu f_T .
\end{equation}
Although specific choices of the function $f$ yield conservative theories, the Herglotz variational principle greatly expands this class of viable conservative models \cite{Chakraborty2013}.

\section{Scalar-tensor Herglotz-type \texorpdfstring{$f(R,T)$}{rep} gravity}\label{sec2}

\subsection{Field equations}

 Although one may formulate $f(R,T)$ gravity directly in the geometric representation, the theory also admits an equivalent scalar-tensor formulation \cite{Lobo_2022, Pinto_2022, Bouali_2023}. Since $f(R,T)$ gravity introduces additional degrees of freedom beyond GR, it is natural to express the theory in this way. Introducing auxiliary scalars $\eta$ and $\zeta$, one rewrites Eq.~(\ref{GeoL}) in the form
\begin{equation}\label{STAux}
\mathcal{L} = f(\eta,\zeta)+(R-\eta)f_\eta+(T-\zeta)f_\zeta+\lambda_\mu s^\mu+16\pi\mathcal{L}_m ,
\end{equation}
leading to equations of motion
\begin{align}\label{matrix}
\begin{pmatrix}
f_{\eta \eta} & f_{\eta\zeta} \\
f_{\zeta\eta} & f_{\zeta\zeta}
\end{pmatrix}\begin{pmatrix}
R-\eta\\
T-\zeta
\end{pmatrix} =0 ,
\end{align}
where subscripts denote differentiation with respect to that variable. For $f_{\eta\eta}f_{\zeta\zeta}\neq f_{\eta\zeta}^2$, the unique nontrivial solution is $R=\eta$ and $T=\zeta$, recovering Eq.~(\ref{GeoL}), and showing equivalence between the two representations.

Now, defining scalar fields $\varphi=f_R$, $\psi=f_T$, and potential $U(\varphi,\psi)=-f+\eta\varphi +\zeta\psi$, Eq.~(\ref{STAux}) may be recast as
\begin{equation}\label{LST}
\mathcal{L} = \varphi R +\psi T -U(\varphi,\psi) +\lambda_\mu s^\mu+16\pi\mathcal{L}_m .
\end{equation}
Varying with respect to $g_{\mu\nu}$ yields
\begin{equation}\label{Fieldeq}
\varphi R_{\mu \nu} - \frac{1}{2} g_{\mu\nu}(\varphi R+\psi T-U) + \left( g_{\mu \nu} \Box - \nabla_{\mu} \nabla_{\nu} \right) \varphi+H_{\mu \nu} =8\pi T_{\mu \nu} - T_{\mu \nu} \psi - \Theta_{\mu \nu} \psi,
\end{equation}
with
\begin{equation}
H_{\mu\nu} = \varphi K_{\mu \nu}+\lambda_\mu \partial_\nu \varphi +\lambda_\nu \partial_\mu \varphi - 2g_{\mu\nu}\lambda_\rho \partial^\rho \varphi .
\end{equation}
Variations of $\varphi$ and $\psi$ give
\begin{align}
R &= U_\varphi ,\label{onshell}\\
T &= U_\psi .\label{onshell2}
\end{align}
Tracing Eq.~(\ref{Fieldeq}) leads to
\begin{equation}\label{trace}
U = \frac{1}{2}\varphi R - \frac{3}{2}\Box\varphi-\frac{1}{2}H +4\pi T +\frac{1}{2}\psi (T-\Theta ) .
\end{equation}
Substituting this into Eq.~(\ref{Fieldeq}) yields the traceless form
\begin{align}\label{tracelessfield}
&\varphi\left(R_{\mu\nu} -\frac{1}{4}Rg_{\mu\nu}\right) + \left(\frac{1}{4}g_{\mu\nu}\Box-\nabla_\mu\nabla_\nu\right)\varphi +H_{\mu\nu}-\frac{1}{4}Hg_{\mu\nu} \nonumber\\
&= 8\pi\left(T_{\mu\nu} -\frac{1}{4}Tg_{\mu\nu}\right)- \psi\left[ T_{\mu\nu}+\Theta_{\mu\nu}-\frac{1}{4}\left( T+\Theta\right)g_{\mu\nu}\right].
\end{align}
Finally, taking the divergence of Eq.~(\ref{Fieldeq}) and using Eqs.~(\ref{onshell})–(\ref{onshell2}) yields
\begin{equation}\label{conservationeq}
(8\pi +\psi)\nabla^\mu T_{\mu\nu} = \nabla^\mu H_{\mu\nu}+\frac{1}{2}\psi\nabla_\nu(2\mathcal{L}_m-T) +(T_{\mu\nu}+\Theta_{\mu\nu})\nabla^\mu \psi .
\end{equation}
Thus, conservation may be retained by imposing
\begin{equation}\label{consconstraint}
\nabla^\mu H_{\mu\nu}=\frac{1}{2}\psi\nabla_\nu(T-2\mathcal{L}_m) -(T_{\mu\nu}+\Theta_{\mu\nu})\nabla^\mu \psi .
\end{equation}

\subsection{Cosmological applications}

After obtaining the field equations, one can explore a cosmological model in the scalar-tensor $f(R,T)$ Herglotz framework by assuming a homogeneous, isotropic, and flat Friedmann-Lemaitre-Robertson-Walker (FLRW) metric 
\begin{align}
    ds^2 = -dt^2 +a^2(t)\delta_{ij}dx^idx^j ,
\end{align}
where $a(t)$ is the scale factor and $t$ is the comoving time. Then, given the spatial homogeneity of the metric, the Herglotz field must take the general form
\begin{align}
    \lambda_\mu = (\phi(t),0,0,0),
\end{align}
$\phi(t)$ being a smooth scalar function of the cosmological time. Similarly, one must have that $f(R,T)$ and its subsequent derivatives only depend on $t$. The matter Lagrangian $\mathcal{L}_m=p$ is taken, leading to the perfect fluid energy-momentum tensor
\begin{align}\label{T}
    T_{\mu\nu} = (p+\rho)u_\mu u_\nu+pg_{\mu\nu} ,
\end{align}
 with pressure $p$ and energy density $\rho$, where the fluid four-velocity obeys $u_\mu u^\mu = -1$.

\subsubsection{The generalized Friedmann equations}

In this setup the generalized FLRW equations take the form 
\begin{align}
\begin{split}\label{FRW1}
        (\phi\varphi-\dot{\varphi})\frac{\dot{a}}{a}-\varphi\left(\frac{\dot{a}}{a}\right)^2 &= \frac{\psi}{6}\left(p-3\rho\right)-\frac{1}{6}U-\frac{8\pi}{3}\rho ,
\end{split}\\\begin{split}\label{FRW2}
     2\varphi\left(\frac{\ddot{a}}{a}-\frac{\dot{a}^2}{a^2}\right) +(\phi \varphi-\dot{\varphi})\frac{\dot{a}}{a} +\ddot{\varphi}-\dot{\phi}\varphi -2\phi\dot{\varphi}+\phi^2\varphi&= -(8\pi+\psi)(\rho+p) .
\end{split}
\end{align}
Additionally, the equations of motion for the scalar fields read
\begin{align}
    U_\varphi &= 6\left(\frac{\ddot{a}}{a}+\frac{\dot{a}^2}{a^2}\right) \label{phieom} , \\
    U_\psi &= 3p-\rho \label{psieom},
\end{align}
for $\varphi$ and $\psi$ respectively. Finally, the conservation equation, namely Eq. (\ref{conservationeq}), in the current cosmological setting takes the form 
\begin{align}\label{FRWconservation}
     &\dot{\rho}+3(\rho+p)\frac{\dot{a}}{a}= \frac{3}{8\pi}\Bigg[-\phi\left(\frac{\ddot{a}}{a}\varphi - \frac{\dot{a}}{a}\dot{\varphi} +\frac{\dot{a}}{a}\phi \varphi\right) -\frac{1}{3} \dot{\psi}(\rho+p) -\psi\left(\frac{1}{2}\dot\rho -\frac{1}{6}\dot{p} +\frac{\dot a}{a}(\rho+p)\right)\Bigg] .
\end{align}

In the cosmology of the scalar-tensor Herglotz type $f(R,T)$ gravity, Eqs. (\ref{FRW1})-(\ref{FRWconservation}) compose a system of five equations where only four are linearly independent. To show this the generalized FLRW equations can be written in an equivalent form
\begin{align}
   3 H^2 &= \frac{8\pi}{\varphi}(\rho+\rho_{eff}) ,\label{Fr1}\\
    2\dot{H} +3H^2 &= -\frac{8\pi}{\varphi}(p+p_{eff}) , \label{Fr2}
\end{align}
using an effective density and pressure,
\begin{align}
    \rho_{eff} &= \frac{1}{8\pi}\left(3(\phi \varphi-\dot{\varphi})H+\frac{\psi}{2}(3\rho-p) +\frac{1}{2}U\right)\label{rhoeff} , \\
    p_{eff} &= \frac{1}{8\pi}\bigg(\ddot{\varphi}-2(\phi \varphi-\dot{\varphi})H -\dot{\phi} \varphi-2\phi \dot{\varphi} +\phi^2 \varphi +\frac{\psi}{2}( 3p-\rho) -\frac{1}{2}U\bigg) \label{peff} ,
\end{align}
with the Hubble function, $H= \dot{a}/a$. 
By eliminating the $3H^2$ term between Eqs.~(\ref{Fr1}) and (\ref{Fr2}) one obtains the equation describing the dynamical evolution of $H$ as
\begin{equation}\label{62}
\dot{H}=-\frac{4\pi}{\varphi}\left(\rho+p+\rho_{eff}+p_{eff}\right).
\end{equation}
Taking the time derivative of Eq.~(\ref{Fr1}), and with the use of Eq.~(\ref{62}), leads to the equation
\begin{align}
\dot{\rho}+3H(\rho+p)&=-\left[\dot{\rho}_{eff}+3H\left(\rho_{eff}+p_{eff}\right)\right]+\left(\rho+\rho_{eff}\right)\frac{\dot{\varphi}}{\varphi}.    
\end{align}
Using Eqs.~(\ref{phieom}), (\ref{psieom}) this is explicitly evaluated as Eq.~(\ref{FRWconservation}). Henceforth, Eq.~(\ref{FRW2}) can be safely discarded so that the system of interest becomes 
\begin{align}
    3H^2 &= \frac{8\pi}{\varphi}\left(\rho +\rho_{eff}\right) \label{cos1},\\
    U_\varphi &= 6(\dot{H}+2H^2) , \label{cos2}\\
    U_\psi &= 3p-\rho ,\label{cos3}\\
    \dot{\rho}+3H(\rho +p) &= -\frac{3}{8\pi}\bigg[ \phi (\varphi\dot{H}+\varphi H^2-\dot{\varphi}H +\phi\varphi H)\nonumber \\
    &+\frac{1}{3}\dot{\psi}\left(\rho+p\right)+\psi\left(\frac{1}{2}\dot\rho-\frac{1}{6}\dot{p} +H(\rho+p)\right) \bigg] .\label{cos4}
\end{align}

To better understand the dynamical evolution, the deceleration parameter $q$, defined as
\begin{equation}
q=\frac{d}{dt}\frac{1}{H}-1  ,
\end{equation}
may be introduced. With the use of Eqs.~(\ref{cos1}) and (\ref{cos2}) the deceleration parameter can be expressed as
\begin{align}
    q&= 1-\frac{\varphi U_\varphi}{16\pi\rho+6(\phi\varphi-\dot{\varphi})H +\psi (3\rho-p) +U }. 
\end{align}
Accelerated expansion corresponds to negative values of $q$, leading to the condition 
\begin{align}
    \varphi U_\varphi > 16\pi \rho + 6\varphi H \phi -6\dot\varphi H +\psi(3\rho -p) +U .
\end{align}

\subsubsection{Specific cosmological model: \texorpdfstring{$U(\varphi,\psi) = \alpha\varphi +\beta\varphi^2 -(1/2\gamma)\psi^2$}{Umod1}} 

 In the following, consider the Universe described by a pressureless fluid (dust), which fixes $p=0$. By imposing an explicit form of the potential 
 \begin{equation}\label{potential}
     U(\varphi,\psi) = \alpha\varphi +\beta\varphi^2 -(1/2\gamma)\psi^2 ,
 \end{equation}
 where $\alpha$, $\beta$ and $\gamma$ are constants with dimensions that render $\varphi$ and $\psi$ dimensionless, and constraining the Herglotz tensor to retain a conservative theory, a closed system is obtained. The latter assumption of demanding energy-momentum conservation closes the system and fixes the density to obey 
 \begin{equation}
     \dot\rho = -3H\rho .
 \end{equation}
 With the following assumptions the generalized FLRW equation reads
 \begin{align}\label{frwp0}
     \dot\varphi = \frac{8\pi}{3H}\rho +\varphi \phi-\varphi H +\frac{1}{2H}\psi\rho +\frac{1}{6H}U,
 \end{align}
 along with the scalar field equations 
 \begin{align}
     U_\varphi &= 6(\dot H +2H^2)\label{Uphi} ,\\
     U_\psi &= -\rho \label{Upsi},
 \end{align}
 and finally the Herglotz field constraint 
 \begin{align} \label{Herconstraint}
     \phi (\varphi\dot{H}+\varphi H^2-\dot{\varphi}H +\phi\varphi H) = -\frac{1}{3}\dot\psi \rho -\frac{1}{2}\psi \dot\rho -\psi H \rho ,
 \end{align}
 which follows from Eq.~(\ref{consconstraint}).
 
To solve the system substitute Eqs.~(\ref{frwp0}) and (\ref{Uphi}) into Eq.~(\ref{Herconstraint}) so the Herglotz field can be formally obtained as
 \begin{align}
     \phi = \frac{2\dot\psi\rho +\psi \dot\rho}{U-\varphi U_\varphi +\rho(16\pi +3\psi)} .
 \end{align}
 Additionally, Eq.~(\ref{Upsi}) immediately gives $\psi =\gamma \rho$. Rescaling the current variables via the present day Hubble constant, $H_0$, one can introduce dimensionless variables ($h,\Phi, r, \tau, \xi, \sigma, \epsilon$) defined by 
\begin{align}
    H&=H_0h,\quad \phi = H_0\Phi,\quad \rho = \frac{3H_0^2}{8\pi}r , \quad \tau=H_0 t ,\\
    \alpha &= 6H_0^2\xi,\,\,\, \beta = 6H_0^2\sigma,\,\,\, \gamma = \frac{256 \pi^2}{15H_0^2}\epsilon ,
\end{align}
which will simplify further analysis. Then the constraint on the Herglotz field reads
\begin{align}
    \Phi = \frac{6\epsilon r \dot{r}}{5r(1+\epsilon r) -5\sigma\varphi^2} ,
\end{align}
so that the relevant system becomes 
\begin{align}
     \frac{dh(\tau)}{d\tau} &= \xi +2\sigma\varphi -2h^2 \label{mod1} , \\
     \frac{d\varphi(\tau)}{d\tau} &=  -\frac{18\epsilon r^2 h \varphi}{5r(1+\epsilon r) -5\sigma\varphi^2} -\varphi h +\frac{\varphi}{h}(\xi +\sigma \varphi) +\frac{r}{h}(1+\epsilon r), \label{mod2}\\
     \frac{d r(\tau)}{d\tau} &= -3h r .
\end{align}

\subsubsection{Numerical results}

Due to the nonlinear character and coupled structure of Eqs.~(\ref{mod1}) and (\ref{mod2}), an analytical solution is not readily accessible. Numerical solutions are therefore obtained and compared with cosmic chronometer data \cite{moresco2020HzTable}. To better facilitate a comparison with late time data, a redshift representation will be used. This leads to the dimensionless matter density,  
\begin{equation}\label{rz}
    r(z) = r_0(1+z)^3 ,
\end{equation}
and the coupled system 
\begin{align}
    -(1+z)h\frac{d\varphi(z)}{dz} &=   \frac{18\epsilon r_0^2(1+z)^5\varphi}{5r_0(1+z)^3(1+\epsilon r_0(1+z)^3) -5\sigma\varphi^2} -\varphi h \nonumber \\&+\frac{\varphi}{h}(\xi +\sigma \varphi) +\frac{r_0(1+z)^3}{h}(1+\epsilon r_0(1+z)^3), \label{zmod1}\\
    -(1+z)h\frac{dh(z)}{dz} &= \xi +2\sigma\varphi -2h^2 ,\label{zmod2}
\end{align}
with initial conditions defined as $r(0)= r_0$, $\varphi(0)=\varphi_0$ and $h(0)=1$. The above equations uniquely determine both the dimensionless Herglotz and $\psi$ field by 
\begin{align}
    \Phi(z) &=  \frac{18\epsilon r_0^2(1+z)^5}{5r_0(1+z)^3(1+\epsilon r_0(1+z)^3) -5\sigma\varphi^2} ,\\
    \psi(z) &= \frac{32 \pi r_0}{5}(1+z)^3 .
\end{align}

The remainder of this section will be to demonstrate that the above model with free parameters $H_0, r_0, \varphi_0, \xi, \sigma, \epsilon$, provides a viable description of the cosmic chronometer data. To accomplish this Eqs.~(\ref{zmod1}) and (\ref{zmod2}) are numerically integrated while fixing $H_0, r_0, \xi, \sigma, \epsilon$ and varying $\varphi_0$. The following parameters will be adopted
\begin{equation}
    H_0=67.36 \,\text{km s}^{-1} \,\text{Mpc}^{-1},\,\, r_0 = 0.3153,\,\,\xi=2, \,\,\sigma=-0.13,\,\, \epsilon=3.7,\,\,\varphi_0 \in \{1.7,1.8,1.9 \} .
\end{equation}
These parameters are not necessarily optimal fits to the data; nevertheless, they are chosen to illustrate a solution that is consistent with observational constraints. A detailed statistical analysis aimed at determining optimal parameter values is left for future work. In addition to cosmic chronometer data comparison, the deceleration parameter
\begin{equation}
    q(z)=(1+z)\frac{1}{h(z)} \frac{dh(z)}{dz}-1,
\end{equation}
and jerk parameter
\begin{equation}
    j(z)=q(z)(2q(z)+1)+(1+z)\frac{dq(z)}{dz} ,
\end{equation}
will be examined. These quantities allow for the analysis of the cosmographic behavior while the $Om(z)$ diagnostic defined by
\begin{equation}
    Om(z)=\frac{h^2(z)-1}{(1+z)^3-1},
\end{equation}
can judge quintessence or phantom-like behavior. All results are compared with the standard $\Lambda$CDM model, characterized by the Hubble function
\begin{equation}
    H(z)=\tensor{H}{_0}\sqrt{\tensor{\Omega}{_m}(1+z)^3+\left(1-\tensor{\Omega}{_m}  \right)},
\end{equation}
with parameters $\tensor{H}{_0}=67.36$ km s$^{-1}$ Mpc$^{-1}$, $\tensor{\Omega}{_m}=0.3153$ \cite{refId0}. 

Additionally, results will be compared to those of a nonconservative and non-Herglotz dependent model with the same potential $U(\varphi,\psi)$. This model was developed in \cite{Pinto_2022} and further statistically optimized in \cite{Bouali_2023}. In particular, Eqs.~(\ref{rz})-(\ref{zmod2}) reduce to Eqs.~(33)-(35) in \cite{Bouali_2023} for a nonconservative energy-momentum tensor and a vanishing Herglotz field.

\begin{figure*}[htbp]
\centering
\includegraphics[width=0.490\linewidth]{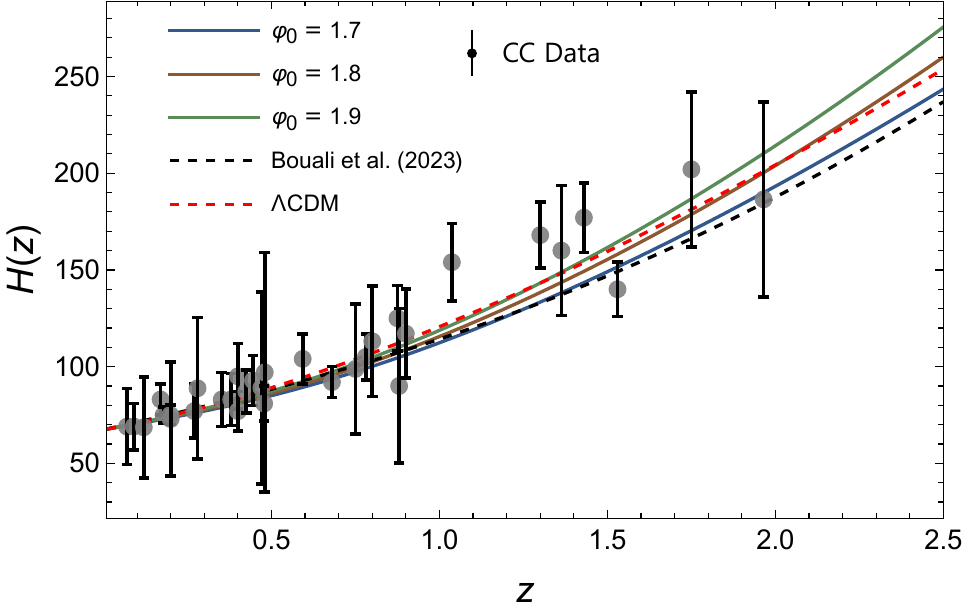} 
\includegraphics[width=0.490\linewidth]{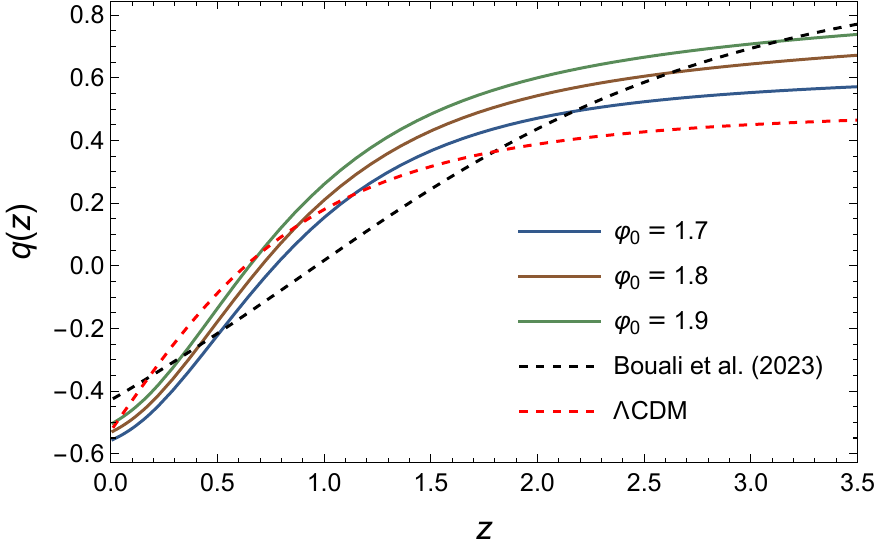}
\caption{The Hubble function (left) and deceleration parameter (right) for initial conditions $\varphi_0 = 1.7$, $\varphi_0 = 1.8$ and $\varphi_0=1.9$ with scalar potential $U(\varphi,\psi) = \alpha\varphi +\beta\varphi^2 -(1/2\gamma)\psi^2 $. Each case is compared with the $\Lambda$CDM model and model I in \cite{Bouali_2023}. Additionally, the Hubble function is compared with cosmic chronometer data.}
\label{fig1}
\end{figure*}

As shown in the left panel of Fig.~\ref{fig1}, for low redshifts (\(z \lesssim 2\)) the model is in good agreement with the cosmic chronometer data. Additionally, the initial condition $\varphi_0=1.8 $ closely matches the Hubble function of the \(\Lambda\)CDM model, whereas $\varphi_0=1.7$ matches closely with \cite{Bouali_2023}\footnote{The statistical analysis used in \cite{Bouali_2023} leads to slightly different values for the Hubble constant $H_0$ and matter density $\Omega_m$ compared to those  used here. In order to make meaningful comparisons, the values $h_0=0.6736$ and $\Omega_m=0.3153$ will be applied to model I in \cite{Bouali_2023}, leaving the behavior unchanged, but producing small translations to some figures found in \cite{Bouali_2023}.}. Initial conditions $\varphi_0\gtrsim 1.9$ lead to faster expansion for larger redshift, while initial conditions $\varphi_0\lesssim 1.7$ lead to slower expansion for larger redshift when compared with \(\Lambda\)CDM. The right panel of Fig.~\ref{fig1} shows that the deceleration parameter follows the global behavior predicted by the $\Lambda$CDM model and \cite{Bouali_2023}. However, for \(z \lesssim 1\) the local behavior has an increasing slope while for \( z  \gtrsim 1\) it becomes decreasing. This same behavior also occurs in \cite{Bouali_2023}, but at a much more moderate rate, while for $\Lambda$CDM the slope is only decreasing. Lower values of $\varphi_0$ tend to delay the transition from increasing to decreasing behavior.

\begin{figure*}[htbp]
\centering
\includegraphics[width=0.490\linewidth]{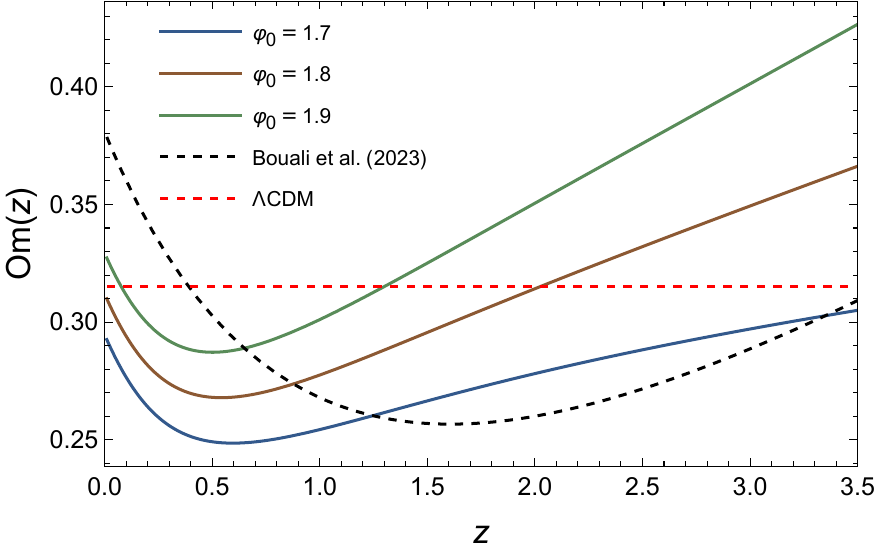} 
\includegraphics[width=0.490\linewidth]{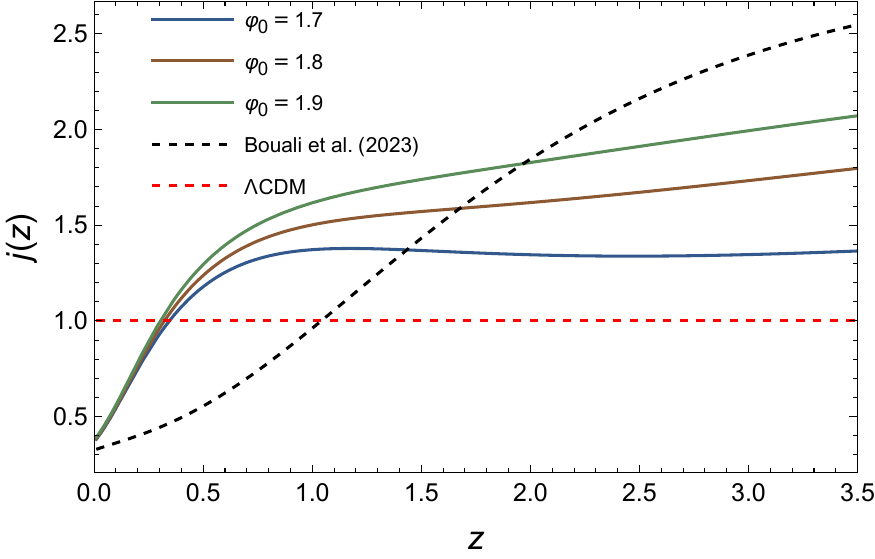}
\caption{The $Om(z)$ diagnostic (left) and jerk parameter (right) for initial conditions $\varphi_0 = 1.7$, $\varphi_0 = 1.8$ and $\varphi_0=1.9$ with scalar potential $U(\varphi,\psi) = \alpha\varphi +\beta\varphi^2 -(1/2\gamma)\psi^2 $.}
\label{fig2}
\end{figure*}
\begin{figure*}[htbp]
\centering
\includegraphics[width=0.490\linewidth]{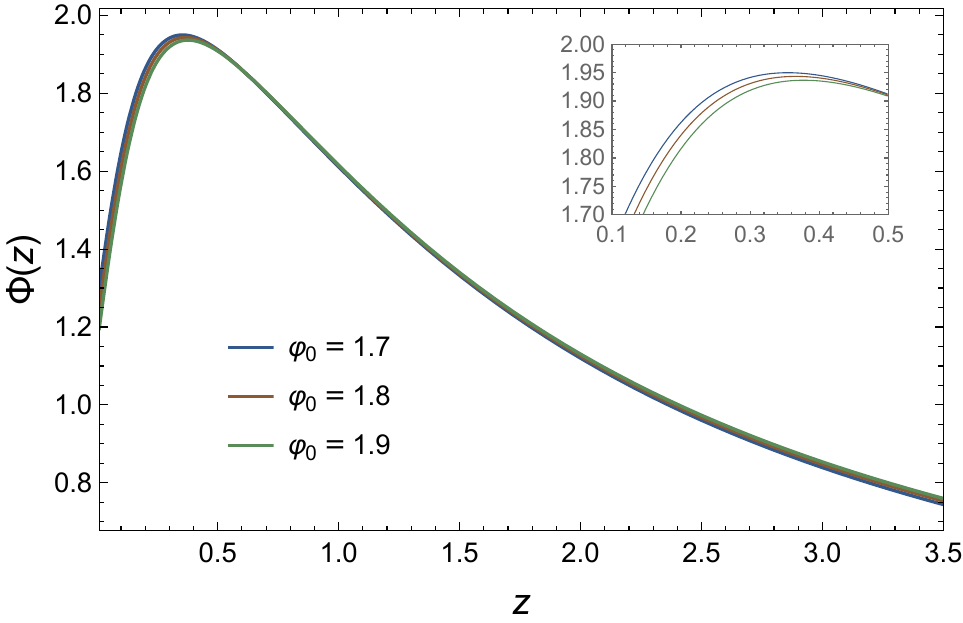}
\includegraphics[width=0.490\linewidth]{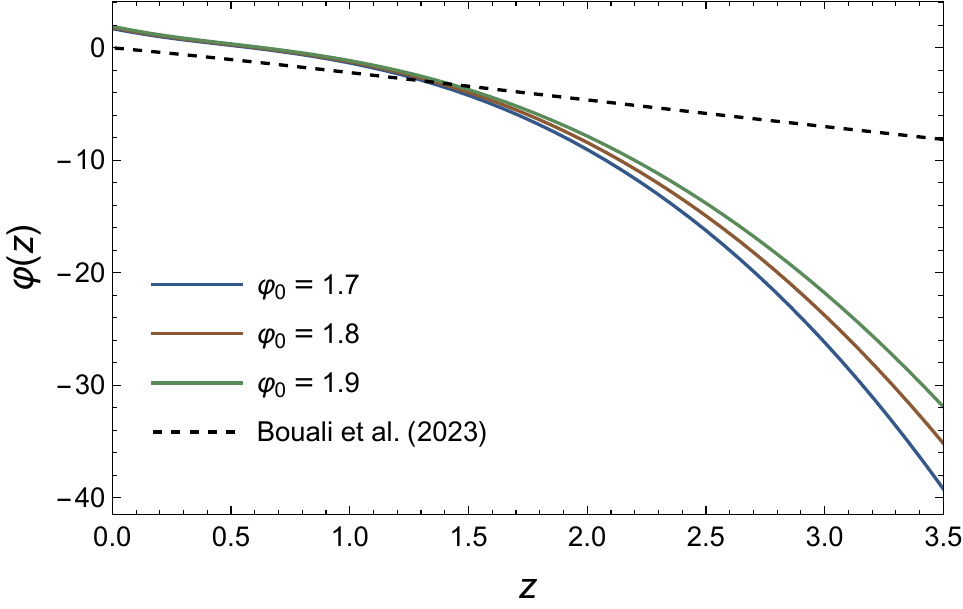} 
\caption{The Herglotz field (left) and scalar field $\varphi$ (right) for initial conditions $\varphi_0 = 1.7$, $\varphi_0 = 1.8$ and $\varphi_0=1.9$ with scalar potential $U(\varphi,\psi) = \alpha\varphi +\beta\varphi^2 -(1/2\gamma)\psi^2 $.}
\label{fig3}
\end{figure*}

The $Om(z)$ diagnostic is shown in the left panel of Fig.~\ref{fig2}. In each initial condition case, $Om(z)$ exhibits a negative slope for $z \lesssim 0.5$, suggesting phantom-like behavior, and a positive slope for $z \gtrsim 0.5$, indicative of quintessence-like behavior. For the $\Lambda$CDM model, it remains constant and equal to the matter density. The right panel of Figure~\ref{fig2} shows the jerk parameter $j(z)$. In $\Lambda$CDM, $j(z) \equiv 1, \forall z \geq 0$, thus any deviations indicate departures from standard cosmology. In the Herglotz-type model, $j(z)$ increases with redshift rapidly until $z\sim 0.5$ where it becomes dependent on the initial condition $\varphi_0$. For values $\varphi_0\lesssim 1.7$ the jerk parameter appears to slowly decay at higher redshift. For values $\varphi_0\gtrsim 1.8$ it appears to slowly increase (with larger $\varphi_0$ indicating faster growth) at higher redshift.

The left panel of Fig.~\ref{fig3} shows the redshift evolution of the Herglotz field $\Phi(z)$. This evolution resembles that of a black-body curve and attains its maximum at $z\sim 0.4$. The initial condition $\varphi_0$ has little effect, but decreasing it tends to increase the Herglotz field peak and shift it  to earlier redshift, essentially playing the role of inverse temperature in the black-body analogy. The right panel of Fig.~\ref{fig3} shows the redshift evolution of the dynamical scalar field $\varphi(z)$. Varying its initial condition leads to larger differences with increasing redshift, while in every case it is strictly decreasing overall.

Comparing the conserved Herglotz-type model with the non-conserved, Herglotz-free model analyzed in \cite{Bouali_2023}, one finds behaviors that are very similar in each figure. The only consistent difference being changes occur less rapidly and at higher redshifts in the non-conserved model. This is most prevalent in the $Om(z)$ diagnostic where both models admit phantom-like behavior followed by quintessence-like behavior, with the non-conserved model showing these behaviors more gradually and at a higher redshift as noted above.

 \section{Discussion and final remarks}\label{conc}

 In the present work, the extension of Herglotz-type $f(R,T)$ gravity was shown to have an equivalent scalar-tensor representation with dynamical fields $\varphi$, $\psi$ and interaction potential $U(\varphi,\psi)$ encoding the extra degrees of freedom absent from General Relativity. Formulating non-minimally coupled gravity theories within scalar-tensor representations becomes of great interest when exploring the dynamical degrees. For $f(R,T)$ gravity in particular, the field $\varphi$ describes the geometric degree of freedom while the field $\psi$ describes the matter degree of freedom. It has been previously shown that these degrees of freedom could lead to particle production/annihilation through the irreversible thermodynamics of open systems \cite{Pinto_2022, Bouali_2023}. This is achieved due to the nonconservation of the energy-momentum tensor present in geometry-matter coupled theories such as $f(R,T)$ gravity. It was also shown in this work (and previously \cite{Wazny_2025}), that the nonconservation persists when using the Herglotz variational principle instead of the typical Lagrangian one. However, in the latter approach the Herglotz vector, although usually general, can be constrained such that the theory preserves its conservation. In this sense the introduction of the Herglotz vector has the ability to essentially cancel the geometry-matter coupling nonconservation while still retaining the dynamical degrees of freedom from the scalar fields.

This behavior was explored in this work through a cosmological model within the scalar-tensor Herglotz-type $f(R,T)$ gravity framework. Using the potential $U(\varphi,\psi)  = \alpha\varphi +\beta\varphi^2 -(1/2\gamma)\psi^2$, the Herglotz vector was constrained such that the theory remained conservative. Although the method of retaining conservation was previously explored in \cite{Lobo_2022} while using an arbitrary potential, it automatically determined the form of the matter scalar $\psi$. In contrast, even when demanding a vanishing energy-momentum tensor, the matter field was free to take the same form as in the nonconservative theory \cite{Pinto_2022} with the same potential used in this work. Additionally, it was found that the dynamics for the Hubble function $H(z)$, deceleration parameter $q(z)$, jerk parameter $j(z)$, scalar field $\varphi(z)$, and $Om(z)$ diagnostic retained the same behavior as \cite{Bouali_2023}. In both interpretations the observational data is described by the model, but with the Herglotz-type model having the advantage of doing so within a conservative or nonconservative cosmology. As such, this creates the opportunity to describe both open and closed systems within the same framework, depending on the structure of the Herglotz vector. For example, if particle production/annihilation occurs on microscopic scales, the Herglotz vector could take a form in this frame ensuring nonconservation, while taking another form on the macroscopic scale retaining conservation. An analogy can be drawn to systems not in equilibrium where a source can lose energy to the surrounding system but the overall energy must remain constant. 

The results obtained in this work strengthen scalar-tensor Herglotz $f(R,T)$ gravity as a promising alternative to the standard $\Lambda$CDM cosmological model. In particular, the scalar-tensor representation reveals that the accelerated expansion of the Universe can naturally emerge from the dynamics of the scalar degree of freedom, thereby eliminating the need to introduce an explicit cosmological constant. This not only avoids the fine-tuning problems typically associated with $\Lambda$, but also offers a dynamical mechanism capable of adapting to different cosmological epochs. While this is true in the standard scalar-tensor $f(R,T)$ formalism, the addition of the Herglotz vector provides the ability to achieve the same results while maintaining a conserved energy-momentum tensor.

Moreover, the predictions of the theory are compatible with current observational data, including the late-time expansion rate. In certain regimes, the model also provides deviations from $\Lambda$CDM that may help address existing tensions, such as those related to the Hubble parameter. Future work may include confronting the theory with large-scale structure data, gravitational-wave propagation constraints, and cosmic microwave background observations, which can further test the viability of the model and potentially distinguish it from the standard cosmological paradigm.

\begin{acknowledgments}
The author thanks both Lehel Csillag and Tiberiu Harko for valuable discussions and improving the quality of the manuscript. The author also thanks Tiago B. Gonçalves for discussions about energy-momentum conservation in the scalar-tensor $f(R,T)$ theory.
\end{acknowledgments}

\bibliography{notes}

\end{document}